\documentclass[10pt]{article}
\usepackage{graphicx,pdflscape}

\newcommand{\beq}{\begin{equation}}
\newcommand{\eeq}{\end{equation}}

\begin{document}
\title{ What Does The Korringa Ratio Measure?}
\author{B. Sriram Shastry$^*$ and Elihu Abrahams$^\dagger$\\
\small \em $^*$AT \& T Bell Laboratories, Murray Hill, New Jersey 07974,\\
\small \em $^\dagger$Serin Physics Laboratory, Rutgers University, PO Box 849,
Piscataway, New Jersey 08855}

\date{24 June 1993}
\maketitle
\begin{abstract}
We present an analysis of the Korringa ratio in a dirty metal,
emphasizing  the case where a Stoner enhancement of the uniform
susceptibilty is present. We find that
the relaxation rates are  significantly enhanced by disorder,
and that the inverse problem of determining
the bare density of states from a study of the change of the Knight
shift and relaxation rates with some parameter, such as pressure,
has rather constrained solutions, with the disorder playing an important
role. Some preliminary applications to the case of chemical substitution in
the Rb$_{3-x}$K$_x $C$_{60}$ family of superconductors is presented and
some other relevant systems are mentioned.

\end{abstract}



\section*{Introduction}

Nuclear magnetic resonance (NMR) is an important experimental tool
for elucidating the physics of electronic materials whose properties
are determined by strong electron-electron correlations.  For
example, NMR data for high-$T_c$ superconductors and for low
dimensional conductors has given us important information about
their electronic properties.  Furthermore, such data are key
benchmarks against which theoretical approaches must be measured.

Basic NMR measurements in a metal involve the magnetic resonance
field shift $K$ and the spin relaxation rate $1/T_1$ of a given
nucleus in the material.\cite{kor} The effects of electron-electron
interactions on these quantities have been studied theoretically for
nearly forty years\cite{dp,toru,seb} and it is often taken for granted that a
deviation of the
``Korringa ratio" ${\cal K} = 1/(T_1TK^2)$ from a constant is a signature
of the
effect of strong correlations.  In particular, the standard canon is that if
the
Korringa ratio exceeds unity, it is evidence for an antiferromagnetic
enhancement of the electron susceptibility away from zero wave vector
which enters $1/T_1$,\cite{seb} while if ${\cal K}$ is less than unity it is
because
there is a (ferromagnetic) Stoner enhancement of the uniform
susceptibility which enters $K$.\cite{dp,toru} It is our purpose to point out
that the
well-known corrections to the susceptibility due to
disorder\cite{fulu} influence the Korringa ratio in an important
way which must be considered when drawing conclusions from the
experimental results. We shall show that disorder significantly enhances
the Korringa ratio and that it must be included when an attempt is made to
estimate Stoner factors from ${\cal K}$. In what follows, we shall not
consider the case of antiferromagnetic enhancements of the wave-vector
dependent electron susceptibility. Our discussion of the ferromagnetic
Stoner enhancement is sufficient to illustrate the point.

When the resonance field shift is determined solely by the local spin
polarization of the
conduction electrons, it is called the Knight shift and it is proportional
to the hyperfine coupling $g \gamma_N A$ and the uniform
conduction electron susceptibility $\chi({\bf q} \rightarrow 0,\,
\omega=0)$.  Here, $\gamma_N$ is the nuclear gyromagnetic
ratio so that $A$ has dimension of magnetic field. At the same time,
$1/T_1$ depends on $A^2$ and on a density of electron
magnetic excitations which is given by the imaginary part of $\chi$.
The latter is averaged over all wave vectors weighted by a form
factor\cite{bss} which depends on the position of the nucleus in the
unit cell and may also include the effect of a transferred hyperfine
interaction.
Thus we have
\beq
K = A\chi(0,0)/\mu_B
\eeq
\beq
1/T_1 = k_B T (\gamma/\mu_B)^2\sum_{{\bf q}}
A^2({\bf q}) \chi''({\bf q},\,\omega_0)/\omega_0, \label{t1}
\eeq
where  $A({\bf q})$ is the ``{\bf q}-dependent hyperfine" coupling
which involves the form factor mentioned above.

The case of free electrons is particularly simple. The susceptibility is
of the Lindhard form:
\beq
\chi'_L(q,\,\omega) = \chi_P[1 - (q/2k_F)^2/3],\;\;
\chi''_L(q,\,\omega) = (\pi \chi_P/2)
(\omega/v_Fq)\,\Theta(v_Fq - \omega), \label{lin}
\eeq
to leading order in
$q,\omega$. Here $\chi_P = m^*k_F\mu_B^2/\pi^2 = 2\rho_0 \mu_B^2$
is the Pauli susceptibility,
$\rho_0$ the one-spin density of states at the Fermi level. With one atom
per unit cell, the form
factor is unity and Eqs.\ (1, 2)
combine to give the well-known Korringa relation:
\beq
1/T_1T = \pi k_B(\gamma_N/\mu_B)^2K^2/4
\eeq
which makes it appear that $1/T_1$ is proportional to the square of
the uniform susceptibility, or to the square of the density of states.
Conclusions have often been drawn from this.
However, this dependence is accidental; from Eq.\ (2) we see that
$1/T_1 \propto \chi/\Gamma$, where
$\Gamma$ is a characteristic energy associated with the
susceptibility at the important range of wave vectors in the sum. In
any case, the Korringa product will cease to be temperature
independent whenever the susceptibility is temperature dependent.

In what follows, we
show that diffusive corrections to $\chi({\bf q},\omega)$ which occur
in a disordered metal\cite{fulu} can also enhance $1/T_1$ and must
be considered when the mean free path is short, as is the case, for
example, in some high-$T_c$ samples and certainly in
M$_3$C$_{60}$.\cite{tyk}

\section*{The Calculation}
We start with the usual RPA result for the susceptibility
$\chi(q,\omega)$
\beq
\chi(q,\omega)^{-1}=\chi_0(q,\omega)^{-1}-U/2
\label{rpa} ,
\eeq
where $U$ is the effective Coulomb interaction
and $\chi_0$ is the dynamical susceptibility, incorporating the
effects of disorder, but not those of interactions. We take units such that
$\hbar
= \mu_B = 1$. We can express $\chi_0$ as a sum over impurity lines,
renormalizing
propagators and vertices to leading order in the impurity scattering rate.
This
gives rise
to the usual diffusion propagator \cite{ramalee} for small
$q,\omega$. We can express
the
result in the form
\beq
\chi_0(q,\omega) \simeq \chi_L(q,\omega) \frac{D_0 q^2}{D_0 q^2-
i\omega},
\eeq
where $\chi_L$ is the Lindhard susceptibility
for the undisordered, non-interacting system. $D_0=\frac{1}{3}
v_F \lambda$ is the diffusion constant
with $\lambda$ the mean free path. To
lowest order in disorder  and
interaction, the mean free path is the same one as enters the
resistivity (for point impurities). However, in general, the spin-triplet
electron-hole
channel mean free path enters the susceptibility, and it can be quite
different (smaller) from the one involved in charge transport in the
interacting system \cite{fulu}. We have constructed Eq.\ (6) so that it
has the correct behavior for both large and small $q$. Thus, it is
correct to
leading order in the disorder, in the limit of low frequencies, and to
$O(q^2)$ in the $q$ dependence. We note the leading behaviour
for small $\omega$
\beq
\chi_0''(q,\omega)/\omega= \rho_0(\frac{\pi}{2 q v_F}+ \frac{1}{D_0
q^2}),  \label{chimlead}
\eeq
This expression is valid for
essentially all the range $q < 2 k_F$, and shows that the
Landau
damping
[the first term in Eq.\ (\ref{chimlead})] ceases to dominate when
the
dimensionless number
$\pi q \lambda/6 <1$. Then the  diffusion mechanism
dominates the
damping.
Also to lowest order, the real part of $\chi$ is
unchanged by disorder and is still given by Eq.\ (\ref{lin}).

In order to calculate the NMR relaxation rate, we need
$\chi''(q,\omega_0)/\omega_0$ which is obtained from Eq.\
(\ref{rpa}) as
\beq
\frac{\chi''(q,\omega_0)}{\omega_0}=
\frac{\chi_0''(q,\omega_0)}{\omega_0} \frac{1}{[1-U
\chi_0'(q,0)/2]^2}. \label{chim}
\eeq
In Eq.\ (\ref{chim}), we neglected a term $[U\chi''_0(q,\omega_0)]^2$
in the denominator since  at the low NMR frequency $\omega_0$, it
vanishes as $\omega_0^2$. The Stoner denominator in Eq.\ (8) is sensitive to
$\chi_0'(q,0)$. Instead of using the approximation of Eq.\ (6) for it, we can
find the result which is valid for all ${\bf q}$ as follows:

At $\omega=0$ there are no diffuson (i.e. vertex) corrections to
$\chi'_0(q,\omega)$ and it is given simply by
\beq
\chi_0({\bf q},0) = - 2\sum_k\frac{n(\bf k+q) - n(\bf
k)}{\epsilon_{\bf k+q}  - \epsilon_{\bf k}}.
\label{bigchi}
\eeq
When the impurity scattering rate $\Gamma = v_F/2\lambda$ is larger
than the temperature (we take $\hbar = 1$),
\beq
n({\bf k}) = \frac{1}{2} - \frac{1}{\pi}\arctan\frac{\epsilon_{\bf
k}}{\Gamma},
\eeq
where the kinetic energy $\epsilon_{\bf k}$ is measured from the
chemical potential.

We perform the angular integrals in Eq.\ (\ref{bigchi}) in two and
three
dimensions and find
\begin{eqnarray}
(2D)~~~~~\chi'_0(q,0) & = & 2\frac{\rho_0}{ t}\int_0^t x\,dx
\frac{ n(x k_F)}{\sqrt{t^2 - x^2}}\\
\nonumber\\
(3D)~~~~~\chi'_0(q,0) & = & \frac{\rho_0}{t}\int_0^{\infty}
x\, dx \;n(x k_F)\ln{\left| \frac{t+x}{t-x}\right| },
\end{eqnarray}
where $x =k/k_F$ and $t = q/2k_F$, $\eta $ is $k_F\lambda$, and
$\rho_0$ is the one-spin Fermi surface density of states in the appropriate
dimension. These may be evaluated in closed form as $q \rightarrow 0$,
where we find $\chi'(0,0)= 2\rho_0 n({\bf k} = 0)$ in 2D and $\chi'(0,0)=
2\rho_0 \sqrt{1+\sqrt{(1+\eta^{-2})}} / \sqrt{2}$ in 3D. These imply that
disorder changes the effective density of states at the Fermi level
by a small amount given by these formulae, and for consistency, we must
use these factors in estimating the enhancement of the susceptibility
or Knight shift.

The result for the NMR relaxation rate is
obtained by using Eqs.\ (7, 11, 12) in Eq.\ (8) and integrating over $q$. The
results are conveniently expressed in terms of the free electron result:
\beq
1/T_1T = \pi k_B(\gamma_N \rho_0 A/2)^2 \cdot {\cal S}(\rho_0
U,  k_F\lambda),
\eeq
where the enhancement factor ${\cal S}(\rho_0 U, k_F\lambda)$ contains
all the effects of interactions and disorder. In terms of the reduced
variables, we have
\begin{eqnarray}
(2D)~~~~~{\cal S}(\rho_0 U, \eta)  & = & \int_{\kappa}^1 dt
\frac{1 + (2/\pi
\eta t)}{[1 -  U \chi'_0(t,0)]^2}\\
\nonumber\\
(3D)~~~~~{\cal S}(\rho_0 U, \eta) & = & 2\int_0^1 tdt
\frac{1 + (3/\pi
\eta t)}{[1 -  U \chi'_0(t,0)]^2}
\end{eqnarray}
In Eq.\ (15), we have introduced the infra-red cutoff $\kappa$ which is
determined by an inelastic scattering (``Thouless") length beyond which
the diffusion ceases. In particular, $\kappa \simeq
1/\sqrt{\eta k_F\lambda_i}$, where $\lambda_i$ is the inelastic mean
free path. These  integrals can be evaluated for $U=0$, where
we find ${\cal S}(\rho_0 U,\eta)= 1 + 6/(\pi\eta)$ in $3D$ and
in $2D$, ${\cal S}(\rho_0 U,\eta)= 1+ \log(k_F^2 \lambda \lambda_{i})
/ (\pi \eta)$.
Note that ${\cal S}$ is in general bigger than unity  due simply to the
enhanced density of states at low energies implied by the
 diffusive character of spin fluctuations in a disordered metal.

A leading approximation for ${\cal S}$ is obtained from Eqs.\ (14,15) by
replacing
$\chi'_0(t,0)$ by, in $3D$, the approximate form $2\rho_0(1 - t^2/3)$ of Eq.\
(3). In $2D$, the same approximation is $\chi'_0(t,0) \simeq 2\rho_0$.
Therefore, the approximate expressions for  the
enhancement factors are
\begin{eqnarray}
(2D)~~~~~{\cal S}(\rho_0 U, \eta)  & = &
\frac{1}{(1-U\rho_0)^2}[1 - \frac{2\ln \kappa}{\pi \eta}]
\label{leading2d}\\
(3D)~~~~~{\cal S}(\rho_0 U, \eta) & = &
\frac{1}{(1-U\rho_0)^2}\left\{
\frac{1}{1 + y^2} + \frac{3}{\pi k_F\lambda} \left[
\frac{1}{1 + y^2} + \frac{1}{y} \arctan y \right] \right\}, \label{leading3d}
\end{eqnarray}
where $y = \sqrt{U\rho_0/[3(1-U\rho_0)]}$.

We can also calculate the dimensionless Korringa ratio ${\cal K}$
(essentially
the ratio of $1/T_1T$ to the square of the Knight shift K ) as
\begin{equation}
{\cal K}=  {\cal S} (\chi_0/\chi)^2.
\end{equation}
This number is also a measure of the interactions and disorder present in a
Fermi system, and is of course  merely an alternative description to
that implied by ${\cal S}$.
However, in some  systems such as  M$_3$C$_{60}$, the Knight shift
is not easy to estimate, since the measured shift has to be apportioned
into the chemical and Knight shifts with large uncertainties in both,
whence we prefer to present both
${\cal K}$ and ${\cal S}$.

\section*{Conclusions}
We show some figures which illustrate the discussion of
this paper. In Fig.\ 1, we show how disorder affects the enhancement
factor ${\cal S}$ in three dimensions. Even when the interaction
(Stoner) enhancement is small, that is $U\rho_0 < 1$, ${\cal S}$ can be
enhanced by a factor 10 when the disorder is sufficiently
great ($k_F\lambda \simeq 1$). Fig.\ 2 shows the same situation in
two-dimensions. The insets in the above two figures show that the approximate
formulas of
Eqs.\ (17, 18) for ${\cal S}$ in three and two dimensions are sufficiently
accurate except at the largest values of the Stoner factor. Since
experiments usually give the Korringa ratio ${\cal K}$ directly, we plot
it in Figs.\ 3, 4 as a function of disorder and interaction strength. In all
the
two-dimensional plots we chose the cutoff $\kappa =
1/\sqrt{20k_F\lambda}$

In Fig.\ 5, we show a result which is potentially useful for a controlled study
in
M$_3$C$_{60}$ where properties appear to depend rather universally on
lattice spacing. For example, in the series M = K$_{3-x}$Rb$_x$ the
variations in $T_c$, susceptibility\cite{add1} and $1/T_1$\cite{tyk} have
been measured. At
present, there seems to be no substantial variation of resistivity with
$x$,\cite{add2,add3}
thus we take $k_F \lambda$ fixed in the plot which shows the change in
$1/T_1T$ relative to that of the uniform susceptibility $\chi$ as the bare
density of states $\rho_0$ is varied. At the present stage, the uncertainties
in the measurements prevent us from making a good inference of the
value of the Stoner enhancement or of the bare density of states, but we
expect that with increased experimental  accuracy, we should be able to exploit
the present
method to infer such quantities, which are of fundamental interest.

As a matter of fact, Knight shift and NMR relaxation rate data for dilute
non-magnetic alloys have been used for many years to study electronic
structure in these materials.\cite{narath} In view of the disorder
enhancement to $1/T_1$ which we have discussed, the systematics of the
behavior with impurity concentration should be included in all such
analyses. More recently, NMR has been used to study the properties of the
rare-earth and actinide heavy electron materials.\cite{asay} For example,
the superconducting compounds U$_{1-x}$Th$_x$Be$_{13}$ are of special
interest because of the dramatic effects caused by the Th doping. There is
a large increase in $1/T_1$ just above $T_c$ as $x$ goes from $0$ to
$0.033$\cite{doug} which is not well-understood but has been attributed
to a density of states effect. Again, it is important to include the effects of
disorder on the relaxation rate, as we have discussed.

A  system that does show the suggested enhancements
 in $1/T_1T$, is the well-studied Si:P \cite{palanen,alloul},
where the observed NMR relaxation rates are approximately three  orders of
magnitude larger than the clean  free electron values. Here there is the
added complication that there are two species of relaxing fluids, the
dirty electron gas as in the present paper, and the singlet pairs formed
by local moments that fluctuate and relax the nuclei in a distinctive
fashion
as described in \cite{ganlee}. The detailed quantitative breakup between
these
terms is not easy to do and is not attempted here. The point we wish to
make
is that the free electron gas component contributes a very large amount
(see Figs.\ 1-2)
to the observed $1/T_1T$ when {\it both} Stoner enhancement  {\it and}
disorder are present.

\section{Acknowledgements}
We thank many generous experimental colleagues who have shared their
data and wisdom with us: A. Hebard, M. Paalanen, T. Palstra, A. Ramirez, R.
Tycko,
R. Walstedt, T.
Imai. This work was supported in part by NSF grant DMR92-21907 (EA).

\begin{figure}
\centering
\includegraphics[width=0.75\textwidth]{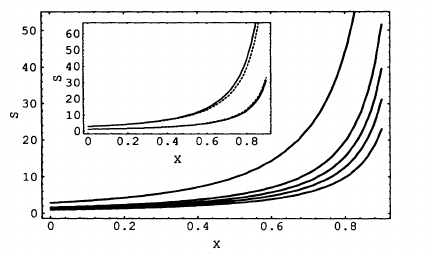}
\caption{${\cal S}$ versus $X = \rho_0 U$ in 3-dimensions for $k_F  \lambda= 1,
3, 5, 10, \infty $
from top to bottom. Inset compares the leading approximation (dashed lines) Eq.
\ (18)
with the exact numerics (solid lines) for $ k_F \lambda= 1, 10$}
\label{fig1}
\end{figure}

\begin{figure}
\centering
\includegraphics[width=0.75\textwidth]{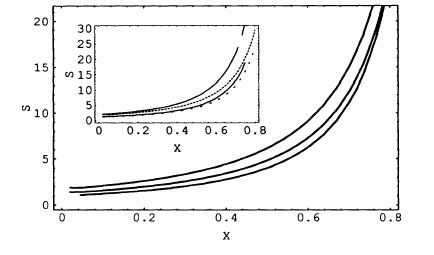}
\caption{${\cal S}$ versus $X =\rho_0 U$ in 2-dimensions for $k_F  \lambda= 1,
3, \infty $
from top to bottom. Inset compares the leading approximation (dashed lines) Eq.
\ (17)
with the exact numerics (solid lines) for $k_F \lambda= 1, 10$
}
\label{fig2}
\end{figure}

\begin{figure}
\centering
\includegraphics[width=0.5\textwidth]{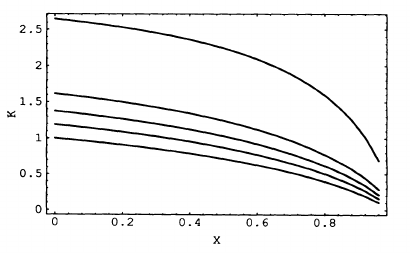}
\caption{The Korringa ratio ${\cal K}$ versus $X=\rho_0 U$ in 3-
dimensions for $k_F \lambda= 1, 3, 5, 10, \infty$
from top to bottom.}
\label{fig3}
\end{figure}

\begin{figure}
\centering
\includegraphics[width=0.7\textwidth]{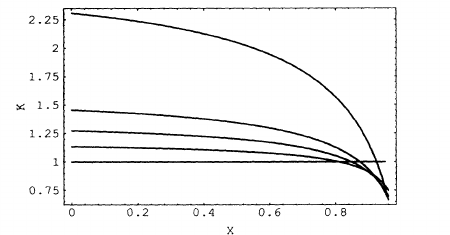}
\caption{The Korringa ratio ${\cal K}$ versus $X =\rho_0 U$ in 2-
dimensions for $k_F \lambda= 1, 3, 5, 10, \infty$
from top to bottom.}
\label{fig4}
\end{figure}

\begin{figure}
\centering
\includegraphics[width=0.7\textwidth]{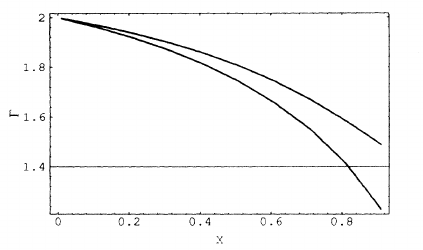}
\caption{ The logarithmic derivative $ \Gamma= d\ln [(\rho_0 U)^2 {\cal
S}(\rho_0 U,k_F \lambda)]/
d \ln [\rho_0 U/(1-\rho_0 U)] $ versus $X=\rho_0 U$ in $3D$. This
corresponds to a variation of the bare density
of states for a fixed $U$ and the resulting change in $1/T_1 T$ relative to
that
in the uniform susceptibility $\chi$ on a logarithmic scale, i.e.
$d \ln (1/T_1T) / d \ln \chi$. Upper (lower) curve is for $k_F \lambda = 1
(\infty)$
}
\label{fig5}
\end{figure}

\end{document}